# On the cause of Zeta potential of a charged vesicle.

## The extra- and intravesicular membrane charges contribute differently to the extra- and intravesicular potential

István P. Sugár

Department of Neurology, Icahn School of Medicine at Mount Sinai, New York, NY 10029

## Abstract

When the charge densities on the inner ($\rho_1$) and outer ($\rho_2$) surface of the vesicle membrane are similar one may assume that the potential at a point closer to the outer/inner surface is more affected by the charges located at the outer/inner surface of the vesicle membrane. However, because of the curvature of the vesicle and the screening effect of the electrolyte ions the situation is more complicated. When $\rho_2 = \rho_1$ then at any finite distance $Z$ from the center of the vesicle the contribution of the charges, located at the outer (or external) membrane surface, to the vesicle potential ($V_{ex}$) is different from the contribution of the charges, located at the inner membrane surface, to the vesicle potential ($V_{in}$). At low electrolyte ion concentrations (from C=0 to 0.00001 [$mol \cdot m^{-3}$]) $V_{ex}/V_{in}>1$ and increases with increasing distance from the center of the vesicle until the external surface of the vesicle membrane. With further increasing distance $V_{ex}/V_{in}$ remains constant at C=0 [$mol \cdot m^{-3}$], while at C>0 [$mol \cdot m^{-3}$] it starts decreasing. However, at high electrolyte ion concentrations (C$\geq$ 0.01 [$mol \cdot m^{-3}$]) with increasing distance from the center of the vesicle $V_{ex}/V_{in}$ reaches its maximum ($V_{ex}/V_{in} \gg 1$) at the internal surface of the vesicle membrane. With further increasing distance, close to the outer surface of the vesicle $V_{ex}/V_{in}$ becomes less than 1 (i.e. the charges at the inner surface of the membrane contribute to the potential more than the charges at the outer surface). Further increasing the distance $V_{ex}/V_{in}$ reaches its minimum and then increases to 1 (i.e. the charges at the inner surface of the membrane contribute to the potential like the charges at the outer surface of the membrane). With increasing electrolyte ion concentration the minimum of $V_{ex}/V_{in}$ is getting deeper (approaching zero) and its location is getting closer to the outer surface of the vesicle. When $\rho_1 \neq \rho_2$ the ratio of the potential produced by charges at the outer membrane surface and the potential produced by charges at the inner membrane surface can be obtained by $(\rho_2/\rho_1) \cdot (V_{ex}/V_{in})$ where $V_{ex}/V_{in}$ is calculated at $\rho_1 = \rho_2$.

## Introduction

The head groups of the lipids of vesicle and cell membranes are neutral or have either electric dipole or single charge. While neutral phospholipids such as sphingomyelin and zwitterionic phosphatidylcholine are located primarily in the outer leaflet of the plasma membrane, most anionic phospholipids, such as phosphatidic acid (PA), phosphatidylserine (PS), phosphatidylethanolamine (PE), and phosphatidylinositol (PI) species, such as phosphatidylinositol 4,5-bisphosphate (PIP$_2$) and phosphatidylinositol (3,4,5)-trisphosphate (PIP$_3$) are mostly located at the inner leaflet [1,2]. It was shown recently that by changing the physical conditions the location of the membrane charges may change [3]. Namely, the zeta potential in hybrid archaeosomes (with 30mol% PLFE (polar lipid fraction E) and 70mol% DPPC (dipalmitoylphosphatidylcholine)) underwent an abrupt increase from -48mV (at 37°C) to -16mV (at 44°C), which was explained by DPPC domain melting and PLFE 'flip-flop' (resulting in a change in the location of the charge of PLFE from one to the other leaflet of the vesicle membrane).

In this paper by using Coulomb's Screened Potential [4-6] and the generalized version of Newton's Shell Theorem [7] we calculate the proportion of the potentials, $V_{ex}/V_{in}$ where $V_{ex}$ and $V_{in}$ is the contribution of the charges located on the outer and inner layer of the membrane to the total potential, respectively. The proportion of these potentials is calculated as a function of the distance from the center of the charged vesicle, $Z$. The intra and extra vesicular character of the $V_{ex}/V_{in}(Z)$ function is different because of the geometrical differences: all the intravesicularly located points are enveloped by the vesicle charges, while none of the points, located extravesicularly, are enveloped by the vesicle charges. However the character of the entire function is independent from charge densities at the outer ($\rho_2$) and inner ($\rho_1$) layer of the membrane since $V_{ex}/V_{in}$ is proportional with $\rho_2/\rho_1$. On the other hand changing the ion concentration of the electrolyte surrounding the vesicle drastically changes the character of the $V_{ex}/V_{in}(Z)$ function.

## Model

In the case of charged vesicle the potential is generated by the charges located at the inner and outer surface of the membrane. The radius of the sphere of the inner and outer membrane surface is $R_1$ and $R_2$, respectively. Since the intra and extra vesicular space is filled with electrolyte in these spaces the potential created by a membrane associated charge, $q$ can be calculated by the screened Coulomb potential [4-6]:

$$V(R) = \frac{k_e q}{\varepsilon R} e^{-R/\lambda_D} \tag{1}$$

where $k_e = (4\pi\varepsilon_0)^{-1}$ is the Coulomb's constant, $\varepsilon_0$ is the electric constant, $\varepsilon$ is the relative static permittivity of the electrolyte, $\lambda_D$ is the Debye length (see Appendix 1) and $R$ is the distance between the charge ($q$) and the considered point, P1.

Figure 1 shows two examples of the screened Coulomb potential, 1) when point P1 is in the intra vesicular space and the charge is in the inner membrane surface (Figure 1A), 2) when point P1 is in the extra vesicular space and the charge is in the outer membrane surface (Figure 1B). By means of the generalized version of Newton's Shell Theorem [Eq.10 in ref.7] the sum of the screened Coulomb potentials created at point P1 (Figure 1A) by the charges at $0 \leq \alpha \leq \pi$ (at the inner membrane surface) is:

$$V_1(Z) = \int_0^\pi \frac{k_e \cdot \rho_1 \cdot 2 \cdot R_1 \cdot \sin(\alpha) \cdot \pi \cdot R_1 \cdot d\alpha}{\varepsilon R(\alpha,Z,R_1)} e^{-R(\alpha,Z,R_1)/\lambda_D} = \frac{k_e \cdot Q_1 \cdot \lambda_D}{\varepsilon \cdot Z \cdot R_1} \cdot e^{-\frac{R_1}{\lambda_D}} \cdot \sinh\left(\frac{Z}{\lambda_D}\right) \quad (2)$$

where $Q_1 = 4R_1^2\pi \cdot \rho_1$ is the total charge of the inner membrane surface, and $\rho_1$ is the charge density at the inner membrane surface. Also, by means of the generalized version of Newton's Shell Theorem [Eq.A2 in ref.7] the sum of the potentials created at point P1 (Figure 1B) by the charges at $0 \leq \alpha \leq \arccos(\frac{Z}{R_2})$ (at the outer membrane surface) is:

$$V_2(Z) = -\frac{k_e \rho_2 2R_2^2 \pi \lambda_D}{\varepsilon Z R_2} [e^w]_{w\{\cos(\alpha)=1\}}^{w\{\cos(\alpha)=\frac{R_2}{Z}\}} = \frac{k_e Q_2 \lambda_D}{2\varepsilon Z R_2}\left[e^{-(Z-R_2)/\lambda_D} - e^{-\sqrt{Z^2-R_2^2}/\lambda_D}\right] \quad (3)$$

where $w = -\sqrt{R_2^2 + Z^2 - 2ZR_2\cos(\alpha)}/\lambda_D$ and $Q_2 = 4R_2^2\pi \cdot \rho_2$ is the total charge of the outer membrane surface, and $\rho_2$ is the charge density at the outer membrane surface.

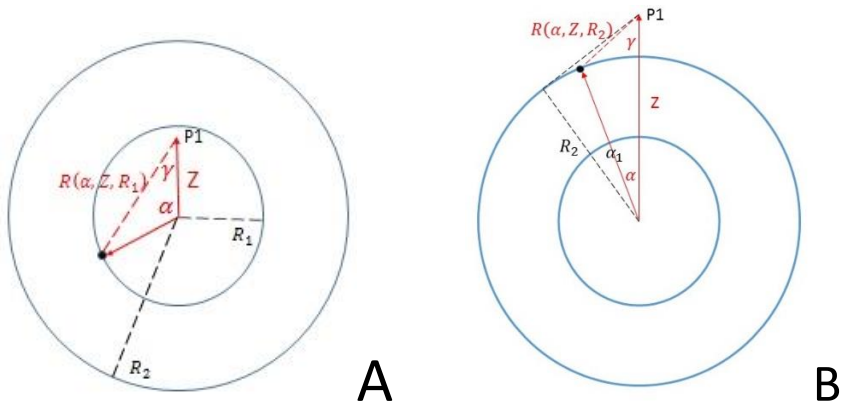

**Figure 1.** *Screened Coulomb potentials*

A) Potential at point P1 (located at the intra vesicular space) created by a charge (marked by black dot, located at the inner surface of the membrane) where $0 \leq \alpha \leq \pi$. The distance between the charge and point P1 is $R(\alpha, Z, R_1) = \sqrt{(R_1 \cdot \sin(\alpha))^2 + (Z - R_1 \cdot \cos(\alpha))^2}$. B) Potential at point P1 (located at the extra vesicular space) created by a charge (marked by black dot, located at the outer surface of the membrane) where $0 \leq \alpha \leq arcos(\frac{R_2}{Z})$. The distance between the charge and point P1 is $R(\alpha, Z, R_2) = \sqrt{(R_2 \cdot \sin(\alpha))^2 + (Z - R_2 \cdot \cos(\alpha))^2}$.

However if the considered point, P1 is located within the membrane, where there is no electrolyte, the potential created by a membrane associated charge, $q$, can be calculated by the regular Coulomb potential [4]:

$$V(r) = \frac{k_e q}{\varepsilon_m R} \tag{4}$$

where $\varepsilon_m (= 2)$ is the relative static permittivity of the membrane. Figure 2 shows two examples of the Coulomb potential, 1) when point P1 is in the membrane and the charge is in the inner membrane surface (Figure 2A), 2) when point P1 is in the membrane and the charge is in the outer membrane surface (Figure 2B). The sum of the potentials created at point P1 (Figure 2A) by the charges at $0 \leq \alpha \leq arcos(\frac{Z}{R_1})$ (at the inner membrane surface) is [from Eq.A6 ref.7]:

$$V_3(Z) = \frac{k_e \cdot Q_1}{2 \cdot \varepsilon_m \cdot R_1 \cdot Z} \cdot \left[ \sqrt{(Z - R_1)(Z + R_1)} - (Z - R_1) \right] \tag{5}$$

The sum of the potentials created at point P1 (Figure 2B) by the charges at $0 \leq \alpha \leq \alpha_1 + \alpha_2$ (at the outer membrane surface) is [from Eq.A7 ref.7]:

$$V_4(Z) = \frac{k_e \cdot Q_2}{2 \cdot \varepsilon_m \cdot R_2 \cdot Z} \cdot \left[ \sqrt{Z^2 + R_2^2 - 2R_1^2 + 2\sqrt{(Z^2 - R_1^2)(R_2^2 - R_1^2)}} - (R_2 - Z) \right] \tag{6}$$

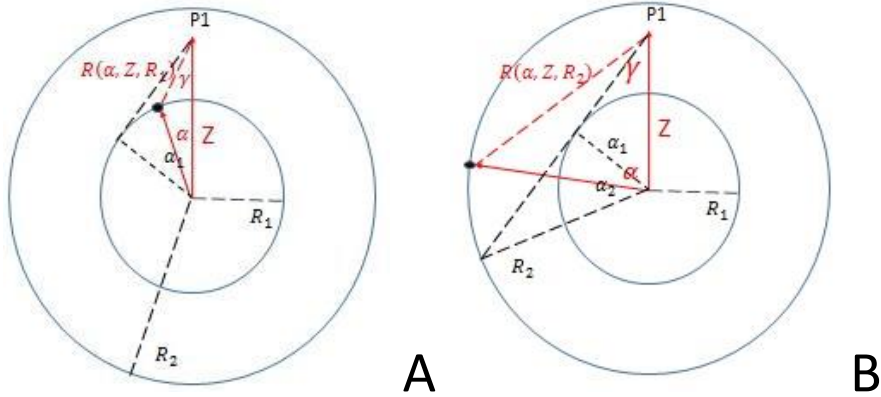

**Figure 2.** *Coulomb potential*

A) Potential at point P1 (located in the membrane) created by a charge (marked by black dot, located at the inner surface of the membrane) where $0 \leq \alpha \leq arcos(\frac{R_1}{Z})$. B) Potential at point P1 (located in the membrane) created by a charge (marked by black dot, located at the outer surface of the membrane) where $0 \leq \alpha \leq \alpha_1 + \alpha_2 = arcos\left(\frac{R_1}{Z}\right) + arcos\left(\frac{R_1}{R_2}\right)$.

Finally, when the point, P1 is in the intra-vesicular electrolyte and the charge, q, is in the outer surface of the membrane (see Figure 3A) or the point is in the extra-vesicular electrolyte and the charge is in either the inner or outer surface of the membrane (see Figure 3C or 3B, respectively) with a distance R, then part of this distance, $R_m$ is in the membrane. In this case the screening length is reduced to $R - R_m$ and the potential of charge q at point P1 is:

$$V(R) = \frac{k_e q}{\varepsilon R} e^{-(R-R_m)/\lambda_D} \tag{7}$$

The sum of the potentials created at point P1 (Figure 3A) by the charges at $0 \leq \alpha \leq \pi$ (at the outer membrane surface) is:

$$V_5(Z) = \frac{k_e Q_2}{2} \int_0^\pi \frac{\sin(\alpha) d\alpha}{\varepsilon R} e^{-\frac{R-R_m}{\lambda_D}} \tag{8}$$

where $R = \sqrt{(R_2 \cdot \sin(\alpha))^2 + (Z - R_2 \cdot \cos(\alpha))^2}$, and $R_m = 0.5 \cdot (r_2 - r_1)$ (where $r_1$ and $r_2$ are defined in Appendix 2).

The sum of the potentials created at point P1 (Figure 3B) by the charges at $arcos(\frac{R_2}{Z}) \leq \alpha \leq \alpha_1 + \alpha_2$ (at the outer membrane surface) is:

$$V_6(Z) = \frac{k_e Q_2}{2} \int_{arcos(\frac{R_2}{Z})}^{\alpha_1+\alpha_2} \frac{\sin(\alpha)d\alpha}{\varepsilon R} e^{-\frac{R-R_m}{\lambda_D}} \tag{9}$$

where $R = \sqrt{(R_2 \cdot \sin(\alpha))^2 + (Z - R_2 \cdot \cos(\alpha))^2}$, and $R_m = r_2$ defined in Appendix 2.

The sum of the potentials created at point P1 (Figure 3C) by the charges at $0 \leq \alpha \leq arcos(\frac{R_1}{Z})$ (at the inner membrane surface) is:

$$V_7(Z) = \frac{k_e Q_1}{2} \int_0^{arcos(\frac{R_1}{Z})} \frac{\sin(\alpha)d\alpha}{\varepsilon R} e^{-\frac{R-R_m}{\lambda_D}} \tag{10}$$

where $= \sqrt{(R_1 \cdot \sin(\alpha))^2 + (Z - R_1 \cdot \cos(\alpha))^2}$, $R_m = 0.5 \cdot (r_2 - r_1)$ ($r_1$ and $r_2$ are defined in Appendix 2).

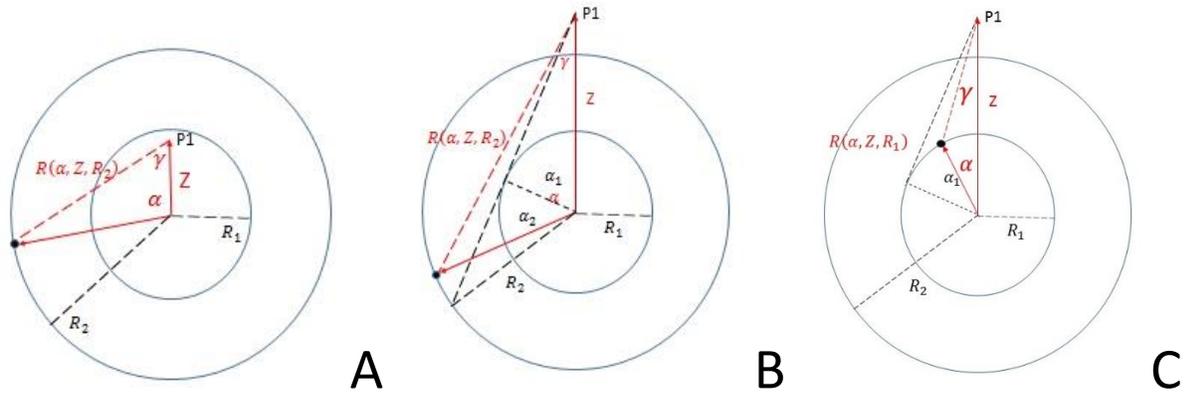

**Figure 3.** *Partially screened Coulomb potentials*

A) Potential at point P1 (located at the intra-vesicular space) created by a charge (marked by black dot, located at the outer surface of the membrane) where $0 \leq \alpha \leq \pi$. B) Potential at point P1 (located at the extra-vesicular space) created by a charge (marked by black dot, located at the outer surface of the membrane) where $arcos\left(\frac{R_2}{Z}\right) \leq \alpha \leq \alpha_1+\alpha_2= arcos\left(\frac{R_1}{Z}\right) + arcos\left(\frac{R_1}{R_2}\right)$. C) Potential at point P1 (located at the extra-vesicular space) created by a charge (marked by black dot, located at the inner surface of the membrane) where $0 \leq \alpha \leq arcos(\frac{R_1}{Z})$.

In Appendix 3 four remaining partially screened Coulomb potentials ($V_8, V_9, V_{10}, V_{11}$) are mentioned. At high electrolyte ion concentrations, because of the strong screening, these potentials are negligible small relative to the above mentioned seven potentials. With these additional four potentials the eleven potentials create the total potential, $V_{tot}(Z)$, of the charged vesicle as follows:

$$V_{tot}(Z) = \begin{cases} V_2 + V_6 + V_7 + V_8 + V_9 \text{ at } Z > R_2 \\ V_3 + V_4 + V_{10} + V_{11} \text{ at } R_1 < Z < R_2 \\ V_1 + V_5 \quad \text{ at } Z < R_1 \end{cases} \quad (11)$$

In ref.7 it was demonstrated that $V_3 + V_4$ closely approaches the total potential within the membrane if the electrolyte concentration (of monovalent ions) is equal or higher than $10\ mol/m^3$.

By using the eleven potentials one can also calculate the proportion of potentials $V_{ex}$ and $V_{in}$ created by the charges located at the outer and inner surface of the membrane, respectively:

$$V_{ex}/V_{in} = \begin{cases} (V_2 + V_6 + V_8)/(V_7 + V_9) \text{ at } Z > R_2 \\ (V_4 + V_{10})/(V_3 + V_{11}) \text{ at } R_1 \leq Z \leq R_2 \\ V_5/V_1 \quad \text{ at } R_1 > Z \end{cases} \quad (12)$$

## Results

Based on Eq.12 the ratio of the potentials created by the charges located at the outer and inner surface of the membrane, $V_{ex}/V_{in}$ has been calculated and plotted against $Z$ (the distance from the center of the charged vesicle; see Figure 4). In these calculations the radius of the vesicle is $R_2 = 10^{-6}[m]$. This radius is the upper limit of LUV's (large unilamellar vesicle), lower limit of GUV's (giant unilamellar vesicle) [8] and a typical size of small prokaryotic cells. The considered membrane thickness is $R_2 - R_1 = 0.005 \cdot 10^{-6}[m]$ [9].

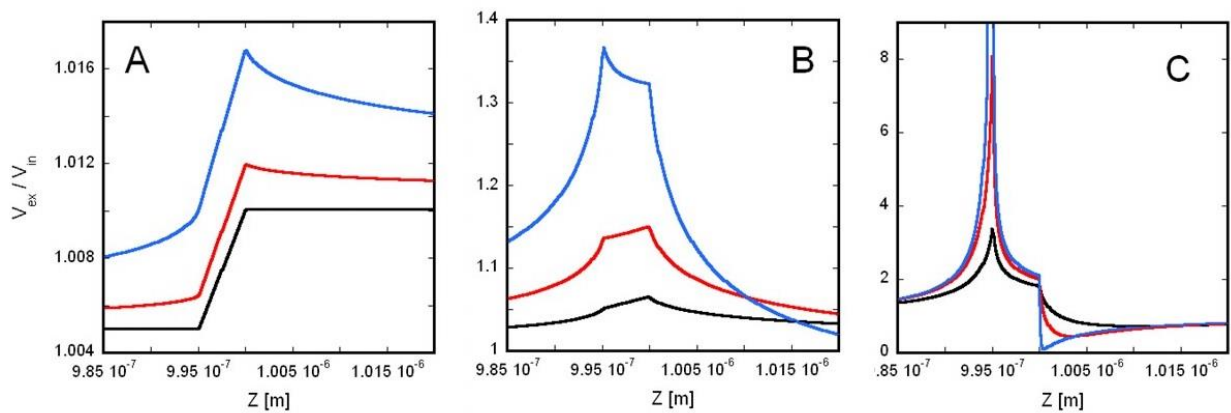

**Figure 4.** *Proportion of the potentials created by the charges located at the outer and inner surface of the vesicle membrane ($V_{ex}/V_{in}$). The electrolyte is electrically neutral both in the intra- and extra-vesicular space. Also the Debye lengths, $\lambda_D$, belonging to the intra- and extra-vesicular space are the same. The inner radius of the vesicle membrane is $R_1 = 0.995 \cdot$*

$10^{-6}[m]$, and the outer radius of the vesicle membrane is $R_2 = 10^{-6}[m]$. The charge densities are the same at the inner and outer surface of the vesicle membrane, i.e. $\rho_2 = \rho_1$. A) Black line: $\lambda_D = \infty [m]$, red line: $\lambda_D = 9.62 \cdot 10^{-6}[m]$, blue line: $\lambda_D = 3.04 \cdot 10^{-6}[m]$    B) Black line: $\lambda_D = 6 \cdot 10^{-7}[m]$, red line: $\lambda_D = 3.04 \cdot 10^{-7}[m]$, blue line: $\lambda_D = 1.5 \cdot 10^{-7}[m]$ C) Black line: $\lambda_D = 3.04 \cdot 10^{-8}[m]$, red line: $\lambda_D = 9.62 \cdot 10^{-9}[m]$, blue line: $\lambda_D = 7.7 \cdot 10^{-10}[m]$.

## Discussion

By using Eq.16 in ref.7 the total potential of a charged vesicle has been calculated as a function of Z. In this paper we are interested in $V_{ex}/V_{in}$, the ratio of the potentials created by the charges located at the outer and inner surface of the vesicle (see Figure 4). As it was mentioned the considered size of the vesicle, $R_2 = 10^{-6}[m]$ is the upper and lower limit of LUV and GUV, respectively and also this size is typical for small prokaryotic cells. Calculating the curves in Figure 4 we assumed the same charge density at the inner and external surface of the vesicle, i.e. $\rho_2 = \rho_1$. However, the shape of the curves do not change when $\rho_2 \neq \rho_1$. This is the case because $V_{ex}$ is proportional to $\rho_2$ and $V_{in}$ is proportional to $\rho_1$ and thus when $\rho_2 \neq \rho_1$ one has to multiply the values at the vertical axis by $\rho_2/\rho_1$. If $Q_2$ and $Q_1$ is the total charge at the outer and inner layer of the membrane, respectively then

$$\frac{\rho_2}{\rho_1} = \frac{Q_2 R_1^2}{Q_1 R_2^2} \tag{13}$$

If $Q_2$ and $Q_1$ are known and also the total potential $V_{tot}$ is known (e.g. by measuring the Zeta potential) one can get $V_{ex}$ and $V_{in}$ by using the ratio $V_r [= V_{ex}/V_{in}]$ (calculated assuming that $\rho_2 = \rho_1$ and plotted in Figure 4) as follows:

$$V_{in} = \frac{V_{tot}}{1 + \left[\frac{Q_2 R_1^2}{Q_1 R_2^2} \cdot V_r\right]} \tag{14}$$

and

$$V_{ex} = V_{tot} \left\{ 1 - \frac{1}{1 + \left[\frac{Q_2 R_1^2}{Q_1 R_2^2} \cdot V_r\right]} \right\} \tag{15}$$

If the charge of the inner layer decreases by a ratio $f$, for example because of the flip-flop of the PLFE molecules [2], the new ratio of the charge densities will be:

$$\frac{\rho_2^{new}}{\rho_1^{new}} = \frac{(Q_2/Q_1) + f}{1 - f} \cdot \frac{R_1^2}{R_2^2} \tag{16}$$

However, if the charge of the outer layer decreases by a ratio $f$ then

$$\frac{\rho_2^{new}}{\rho_1^{new}} = \frac{1-f}{(Q_1/Q_2)+f} \cdot \frac{R_1^2}{R_2^2} \qquad (17)$$

Also, calculating the curves in Figure 4 we took the same Debye length in the intra- and extra-vesicular space. Assuming that the extra- and intra-vesicular space filled by monovalent anions and cations and by using Eq.A1 Table 1 has been created (showing the ion concentration at each Debye length considered in Figure 4).

**Table 1.** *Monovalent ion concentrations of electrolytes and the respective Debye lengths*

| Electrolyte concentration $(mol \cdot m^{-3})$ | Debye length, $\lambda_D$ $(m)$ |
|---|---|
| 0.0 | ∞ |
| 0.000001 | $9.62 \cdot 10^{-6}$ |
| 0.00001 | $3.04 \cdot 10^{-6}$ |
| 0.000257 | $6.00 \cdot 10^{-7}$ |
| 0.001 | $3.04 \cdot 10^{-7}$ |
| 0.00411 | $1.5 \cdot 10^{-7}$ |
| 0.1 | $3.04 \cdot 10^{-8}$ |
| 1.0 | $9.62 \cdot 10^{-9}$ |
| See Appendix 1 | $7.7 \cdot 10^{-10}$ |

Note: the last Debye length in the legend of Figure 4 and in Table 1 is characteristic to the intra- and extra-cellular space of mammalian cells (see Appendix 1).

When the electrolyte ion concentration is zero one can calculate analytically $V_{ex}/V_{in}$ (see black curve in Figure 4A). By means of the Shell Theorem [10] the total intra-, extra-vesicular and intra-membrane potential is:

$$V_{tot}(Z) = \begin{cases} \frac{k_e \rho_2 4\pi R_2^2}{\varepsilon Z} + \frac{k_e \rho_1 4\pi R_1^2}{\varepsilon Z} & Z > R_2 \\ \frac{k_e \rho_2 4\pi R_2^2}{\varepsilon_m R_2} + \frac{k_e \rho_1 4\pi R_1^2}{\varepsilon_m Z} & R_1 < Z < R_2 \\ \frac{k_e \rho_2 4\pi R_2^2}{\varepsilon R_2} + \frac{k_e \rho_1 4\pi R_1^2}{\varepsilon R_1} & Z < R_1 \end{cases}$$

(18)

and dividing the first with the second term and taking $\rho_2 = \rho_1$ we get

$$V_r(Z) = \begin{cases} (R_2/R_1)^2 & Z > R_2 \\ (R_2 Z/R_1^2) & R_1 < Z < R_2 \\ (R_2/R_1) & Z < R_1 \end{cases} \tag{19}$$

At the center of the charged vesicle one can calculate $V_{ex}/V_{in}$ at any Debye length. Since the distance of each charge at the inner (or outer) surface of the vesicle has the same distance from the center thus:

$$\frac{V_{ex}}{V_{in}}(Z=0) = \frac{\frac{k_e \rho_2 4\pi R_2^2}{\varepsilon R_2} e^{-(R_2 - R_m)/\lambda_D}}{\frac{k_e \rho_1 4\pi R_1^2}{\varepsilon R_1} e^{-R_1/\lambda_D}} = \frac{\frac{k_e \rho_2 4\pi R_2^2}{\varepsilon R_2} e^{-R_1/\lambda_D}}{\frac{k_e \rho_1 4\pi R_1^2}{\varepsilon R_1} e^{-R_1/\lambda_D}} = \frac{\rho_2 R_2}{\rho_1 R_1} \tag{20}$$

where the length of the unscreened region $R_m$ is equal with the membrane thickness, i.e. $R_m = R_2 - R_1$.

At high electrolyte concentrations (above $0.004 \ mol \cdot m^{-3}$) $V_{ex}/V_{in}$ has maximum at $Z = R_1$ (see Figure 4B,C). This maximum and its surrounding can be approximated analytically:

$$\frac{V_{ex}}{V_{in}}(R_1 \leq Z) \cong \frac{V_4}{V_3 + [V_{11}(Z = R_1)]}$$

$$= \frac{\frac{k_e \cdot Q_2}{2 \cdot \varepsilon_m \cdot R_2 \cdot Z} \cdot \left[ \sqrt{Z^2 + R_2^2 - 2R_1^2 + 2\sqrt{(Z^2 - R_1^2)(R_2^2 - R_1^2)}} - (R_2 - Z) \right]}{\frac{k_e \cdot Q_1}{2 \cdot \varepsilon_m \cdot R_1 \cdot Z} \left[ \sqrt{(Z - R_1)(Z + R_1)} - (Z - R_1) \right] + \frac{k_e \cdot Q_1 \cdot \lambda_D}{\varepsilon_m \cdot R_1^2} \cdot e^{-\frac{R_1}{\lambda_D}} \cdot \sinh\left(\frac{R_1}{\lambda_D}\right)} \tag{21}$$

where $V_{11}(Z = R_1) = [\varepsilon/\varepsilon_m] \cdot V_1(Z = R_1)$. Note, without including $V_{11}(Z = R_1)$ into Eq.22 $V_{ex}/V_{in}$ would be infinite at $Z = R_1$.

Also at high electrolyte concentrations (above $0.004 \ mol \cdot m^{-3}$) $V_r$ has a minimum at $Z > R_2$ (see Figure 4C). With increasing electrolyte ion concentration the minimum is deeper and getting closer to $R_2 = 10^{-6}[m]$ (see Table 2).

**Table 2**. *Location and the depth of the minimum of $V_r$*

| Debye length, $\lambda_D$ (m) | $Z_{min}$ [m] | $V_r$ at $Z_{min}$ |
|---|---|---|
| $3.04 \cdot 10^{-7}$ | $2.0318 \cdot 10^{-6}$ | 0.9965977 |
| $1.5 \cdot 10^{-7}$ | $1.1167 \cdot 10^{-6}$ | 0.9713244 |
| $3.04 \cdot 10^{-8}$ | $1.0104 \cdot 10^{-6}$ | 0.7302959 |
| $9.62 \cdot 10^{-9}$ | $1.0033 \cdot 10^{-6}$ | 0.4531 |
| $7.7 \cdot 10^{-10}$ | $1.0004 \cdot 10^{-6}$ | 0.102005 |

After the minimum is reached $V_r$ is increasing and approaches one. On the other hand at low electrolyte ion concentrations with increasing distance from the vesicle $V_r$ continuously decreases and approaches one. However, at zero electrolyte ion concentration there is no screening and $V_r[=(R_2/R_1)^2]$ is constant at $Z \geq R_2$ (see Eq.19).

At high electrolyte ion concentrations from Eq.12 at $Z \geq R_2$ only the following three potentials contribute significantly to $V_{ex}/V_{in}$: $V_2$, $V_6$ and $V_7$. Because of the additional strong screening by the intra-vesicular electrolyte the remaining potentials, $V_8$ and $V_9$, are much smaller than $V_2$, $V_6$ and $V_7$. Thus

$$\frac{V_{ex}}{V_{in}} \cong \frac{V_2 + V_6}{V_7} \qquad at \ Z \geq R_2 \tag{22}$$

The reason of the minimum of $V_{ex}/V_{in}$ at high electrolyte ion concentration (see Fig.4C at $Z \geq R_2$) is that the characteristics of function $V_2$ is different from $V_6$ and $V_7$ (see Figs.5A,B). As a consequence $V_2/V_7$ is a continuously increasing function while $V_6/V_7$ is a continuously decreasing function and their sum has a minimum. In Figure 5C the minimum of the dotted blue curve is at $Z = 1.0037 \cdot 10^{-6}[m]$ and its value is 0.45847. These characteristic values of the minimum are close to the respective value in Table 2 listed at $\lambda_D = 9.62 \cdot 10^{-9}[m]$.

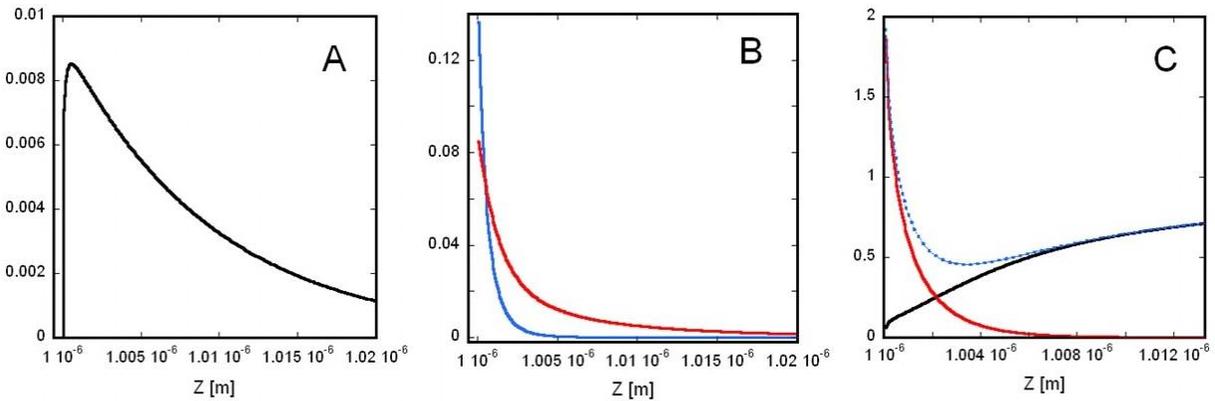

**Figure 5.** *The reason of the minimum of $V_{ex}/V_{in}$ at high electrolyte ion concentration*
A) Black curve: $V_2$ vs. $Z$; B) Blue curve: $V_6$ vs. $Z$, red curve: $V_7$ vs. $Z$; C) Black curve: $V_2/V_7$ vs. $Z$, red curve: $V_6/V_7$ vs. $Z$, blue dotted curve: $(V_2 + V_6)/V_7$ vs. $Z$. The curves were calculated at a Debye length: $\lambda_D = 9.62 \cdot 10^{-9}[m]$ and at $\rho_2 = \rho_1$.

## Conclusions

The contribution of the charges, located at the outer (or external) membrane surface, to the vesicle potential ($V_{ex}$) is different from the contribution of the charges, located at the inner membrane surface, to the vesicle potential ($V_{in}$). $V_{ex}/V_{in}$ is calculated assuming the same charge density at the inner and outer surface of the vesicle membrane, i.e. $\rho_1 = \rho_2$. At low electrolyte ion concentrations (from C=0 to 0.00001 $[mol \cdot m^{-3}]$) $V_{ex}/V_{in}$>1 and increases with increasing distance $Z$ from the center of the vesicle until the outer surface of the vesicle membrane. With further increasing distance $V_{ex}/V_{in}$ remains constant at C=0 $[mol \cdot m^{-3}]$, while at C>0 $[mol \cdot m^{-3}]$ it starts decreasing. However, at high electrolyte ion concentrations (C≥ 0.01 $[mol \cdot m^{-3}]$) with increasing distance from the center of the vesicle $V_{ex}/V_{in}$ reaches its maximum ($V_{ex}/V_{in} \gg 1$) at the inner surface of the vesicle membrane, i.e. at $Z = R_1$. With further increasing distance, close to the outer surface of the vesicle (at $Z \geq R_2$) $V_{ex}/V_{in}$ becomes less than 1 (i.e. the charges at the inner surface of the membrane contribute to the potential more than the charges at the outer surface). Further increasing the distance $V_{ex}/V_{in}$ reaches its minimum and then increases until it becomes equal to 1 at $Z = \infty$ (i.e. the charges at the inner surface of the membrane contribute to the potential like the charges at the outer surface of the membrane). With increasing electrolyte ion concentration the minimum of $V_{ex}/V_{in}$ is getting deeper (approaching zero) and its location is getting closer to the outer surface of the vesicle. If $\rho_1 \neq \rho_2$ the ratio of the potential produced by charges at the outer membrane surface and the potential produced by charges at the inner membrane surface can be obtained by $(\rho_2/\rho_1) \cdot (V_{ex}/V_{in})$ where $V_{ex}/V_{in}$ is calculated at $\rho_1 = \rho_2$.

## Acknowledgement

The author is very thankful for Chinmoy Kumar Ghose.

## Appendix 1

The Debye length in an electrolyte is calculated by [11]

$$\lambda_D = \left(\frac{\varepsilon_0 \varepsilon k_B T}{e^2 N_a \sum_{j=1}^{N} c_j^0 q_j^2}\right)^{\frac{1}{2}} \tag{A1}$$

where $\varepsilon_0 = 8.85 \cdot 10^{-12} [C^2 J^{-1} m^{-1}]$ is the vacuum permittivity, $\varepsilon$ is the relative static permittivity of the electrolyte, $k_B = 1.38 \cdot 10^{-23} [J K^{-1}]$ is the Boltzmann constant, $T[K]$ is the absolute temperature, $e = 1.6 \cdot 10^{-19} [C]$ is the charge of a positive monovalent ion, $N_a = 6 \cdot 10^{23} [mol^{-1}]$ is the Avogadro's number, $c_j^0 \, [mol \cdot m^{-3}]$ is the mean concentration of the j-th species of ions in the electrolyte, $q_j$ is the number of elementary charges in an ion of the j-th species (e.g.in the case of bivalent ions $q_j = 2$). In our calculation we take always $T = 300[K]$ and $\varepsilon = 78$. We consider three types of electrolytes. The first contains only a monovalent cation and a monovalent anion each with concentration $100 \, mol \cdot m^{-3}$. In the case of this electrolyte $\lambda_D = 3.04 \cdot 10^{-9} [m]$. The second models the electrolyte within a mammalian cell containing [12]: 139 $[mol \cdot m^{-3}] \, K^+$; 12 $[mol \cdot m^{-3}] \, Na^+$; 4 $[mol \cdot m^{-3}] \, Cl^-$; 12 $[mol \cdot m^{-3}] \, HCO_3^-$; 138 $[mol \cdot m^{-3}] \, X^-$; 0.8 $[mol \cdot m^{-3}] \, Mg^{2+}$; <0.0002 $[mol \cdot m^{-3}] \, Ca^{2+}$ where $X^-$ represents proteins, which have a net negative charge at the neutral *pH* of blood and cells. In the case of this electrolyte $\lambda_D = 7.7 \cdot 10^{-10} [m]$. The third models the electrolyte at the extracellular space of a mammalian cell containing: 4 $[mol \cdot m^{-3}] \, K^+$; 145 $[mol \cdot m^{-3}] \, Na^+$; 116 $[mol \cdot m^{-3}] \, Cl^-$; 29 $[mol \cdot m^{-3}] \, HCO_3^-$; 9 $[mol \cdot m^{-3}] \, X^-$; 1.5 $[mol \cdot m^{-3}] \, Mg^{2+}$; 1.8 $[mol \cdot m^{-3}] \, Ca^{2+}$. In the case of this electrolyte $\lambda_D = 7.67 \cdot 10^{-10} [m]$.

Note, that the relative static permittivity of the electrolyte decreases with increasing ion concentration. However in the above mentioned concentration region the decrease is within one percent [13,14]. Thus in our calculations the relative static permittivity of the electrolyte is taken $\varepsilon = 78$ at the above electrolyte concentrations.

## Appendix 2 Calculating the length of the cords

First we consider the case when the charge is located on the circle of radius $R_2$ (see black dot on the circle in Figure A1). Here we calculate the length of the chord, $r_2$ shown in Figure A1. The equation of the red line, connecting the black dot with the point at a distance Z from the center of the sphere, is:

$$y = tg(\gamma) \cdot (x + Z) \tag{A2}$$

and the equation of the circle of radius $R_2$ is:

$$x^2 + y^2 = R_2^2 \tag{A3}$$

The solutions of Eqs.A2,A3 results in the coordinates of the two intersections between the straight line and the circle. The solutions are:

$$x_{2/1} = \frac{-Z \cdot tg^2(\gamma) \pm \sqrt{R_2^2 \cdot [tg^2(\gamma)+1] - tg^2(\gamma) \cdot Z^2}}{tg^2(\gamma)+1} \tag{A4}$$

$$y_{2/1} = tg(\gamma) \cdot [x_{2/1} + Z] \tag{A5}$$

From Eqs.A4,A5 we get the length of the chord:

$$r_2 = \sqrt{(x_2 - x_1)^2 + (y_2 - y_1)^2} = 2 \cdot \sqrt{\frac{R_2^2 \cdot [tg^2(\gamma)+1] - tg^2(\gamma) \cdot Z^2}{tg^2(\gamma)+1}} \tag{A6}$$

and based on Figure A1 $tg(\gamma)$ depends on $\alpha$ as follows:

$$tg(\gamma) = \frac{R_2 \cdot \sin(\alpha)}{Z - R_2 \cdot \cos(\alpha)} \tag{A7}$$

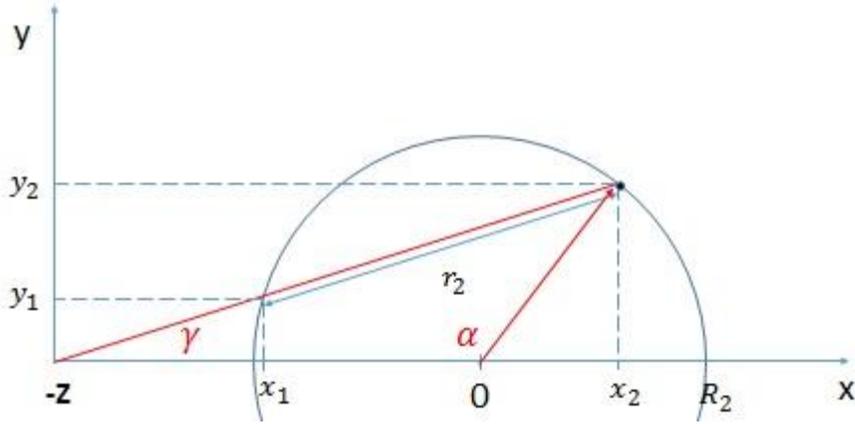

**Figure A1.** *Calculating the length of the chord, $r_2$*

If the straight line intersects the circle with radius $R_1$ the length of the chord, $r_1$ is:

$$r_1 = 2 \cdot \sqrt{\frac{R_1^2 \cdot [tg^2(\gamma)+1] - tg^2(\gamma) \cdot Z^2}{tg^2(\gamma)+1}} \tag{A8}$$

However, if the considered charge is located on the sphere of radius $R_1$ then $tg(\gamma)$ depends on $\alpha$ as follows:

$$tg(\gamma) = \frac{R_1 \cdot \sin(\alpha)}{Z - R_1 \cdot \cos(\alpha)} \tag{A9}$$

In this case calculating the length of the chords $r_1$ and $r_2$ one has to substitute this $tg(\gamma)$ into A8 and A6, respectively.

## Appendix 3 Four more partially screened Coulomb potentials

In Figure A2 four additional partially screened Coulomb potentials are shown.

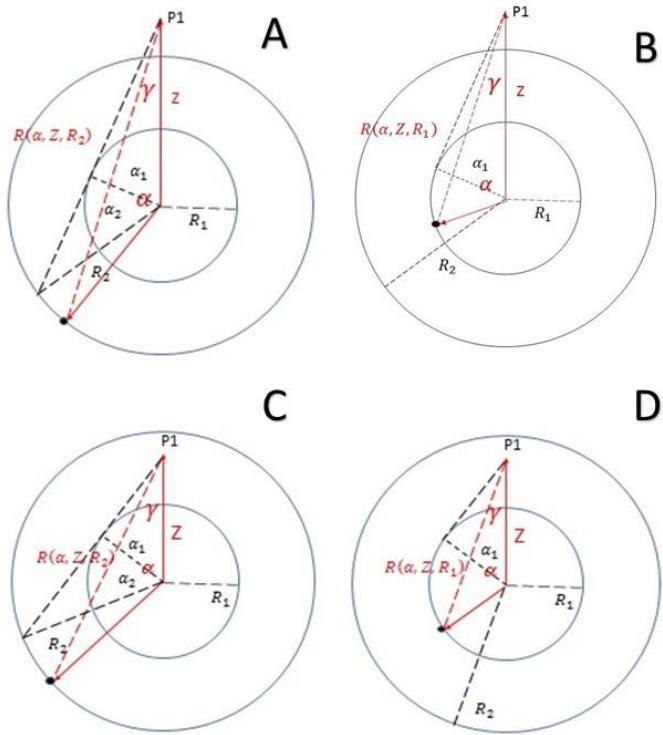

**Figure A2.** *Partially screened Coulomb potentials*

A) Potential at point P1 (located at the extra-vesicular space) created by a charge (marked by black dot, located at the outer surface of the membrane) where $\alpha_1 + \alpha_2 \; [= arcos\left(\frac{R_1}{Z}\right) + arcos\left(\frac{R_1}{R_2}\right)] \leq \alpha \leq \pi$. B) Potential at point P1 (located at the extra-vesicular space) created by a charge (marked by black dot, located at the inner surface of the membrane) where $arcos\left(\frac{R_1}{Z}\right) \leq \alpha \leq \pi$. C) Potential at point P1 (located in the membrane) created by a charge (marked by black dot, located at the outer surface of the membrane) where $\alpha_1 + \alpha_2 \; [= arcos\left(\frac{R_1}{Z}\right) + arcos\left(\frac{R_1}{R_2}\right)] \leq \alpha \leq \pi$. D) Potential at point P1 (located in the membrane) created by a charge (marked by black dot, located at the inner surface of the membrane) where $arcos\left(\frac{R_1}{Z}\right) \leq \alpha \leq \pi$.

The sum of the potentials created at point P1 (Figure A2(A)) by the charges at $\alpha_1 + \alpha_2 \; [= arcos\left(\frac{R_1}{Z}\right) + arcos\left(\frac{R_1}{R_2}\right)] \leq \alpha \leq \pi$ (located at the outer membrane surface) is:

$$V_8(Z) = \frac{k_e Q_2}{2} \int_{\alpha_1+\alpha_2}^{\pi} \frac{\sin(\alpha)\, d\alpha}{\varepsilon R} e^{-\frac{R-R_m}{\lambda_D}}$$

where $R = \sqrt{(R_2 \cdot \sin(\alpha))^2 + (Z - R_2 \cdot \cos(\alpha))^2}$, and $R_m = (r_2 - r_1)$ (where $r_1$ and $r_2$ are defined in Appendix 2).

The sum of the potentials created at point P1 (Figure A2(B)) by the charges at $\alpha_1 [= arcos\left(\frac{R_1}{Z}\right)] \leq \alpha \leq \pi$ (located at the inner membrane surface) is:

$$V_9(Z) = \frac{k_e Q_1}{2} \int_{\alpha_1}^{\pi} \frac{\sin(\alpha)\, d\alpha}{\varepsilon R} e^{-\frac{R-R_m}{\lambda_D}}$$

where $R = \sqrt{(R_1 \cdot \sin(\alpha))^2 + (Z - R_1 \cdot \cos(\alpha))^2}$, and $R_m = 0.5 \cdot (r_2 - r_1)$ ($r_1$ and $r_2$ are defined in Appendix 2).

The sum of the potentials created at point P1 (Figure A2(C)) by the charges at $\alpha_1 + \alpha_2 [= arcos\left(\frac{R_1}{Z}\right) + arcos\left(\frac{R_1}{R_2}\right)] \leq \alpha \leq \pi$ (located at the outer membrane surface) is:

$$V_{10}(Z) = \frac{k_e Q_2}{2} \int_{\alpha_1+\alpha_2}^{\pi} \frac{\sin(\alpha)\, d\alpha}{\varepsilon_m R} e^{-\frac{R-R_m}{\lambda_D}}$$

where $R = \sqrt{(R_2 \cdot \sin(\alpha))^2 + (Z - R_2 \cdot \cos(\alpha))^2}$ and the screening length is $R - R_m = r_1$ ($r_1$ is defined in Appendix 2).

The sum of the potentials created at point P1 (Figure A2(D)) by the charges at $\alpha_1 [= arcos\left(\frac{R_1}{Z}\right)] \leq \alpha \leq \pi$ (located at the inner membrane surface) is:

$$V_{11}(Z) = \frac{k_e Q_1}{2} \int_{\alpha_1}^{\pi} \frac{\sin(\alpha)\, d\alpha}{\varepsilon_m R} e^{-\frac{R-R_m}{\lambda_D}}$$

where $R = \sqrt{(R_1 \cdot \sin(\alpha))^2 + (Z - R_1 \cdot \cos(\alpha))^2}$ and the screening length is $R - R_m = r_1$ (where $r_1$ is defined in Appendix 2).